\documentclass[prl,twocolumn,superscriptaddress,showpacs]{revtex4}
\usepackage{amsmath}
\usepackage{amssymb}
\usepackage{graphicx}
\usepackage{pst-all}

\begin{document}

\title{Hierarchical Nature of the Quantum Hall Effects}
\author{Parsa Bonderson}
\affiliation{Station Q, Microsoft Research, Santa Barbara, California 93106-6105, USA}
\date{\today}

\begin{abstract}
I demonstrate that the wavefunction for a $\nu = n+ \tilde{\nu}$ quantum Hall state with Landau levels $0, 1, \ldots, n-1$ filled and a filling fraction $\tilde{\nu}$ quantum Hall state with $0 < \tilde{\nu} \leq 1$ in the $n$th Landau level can be obtained hierarchically from the $\nu = n$ state by introducing quasielectrons which are then projected into the (conjugate of the) $\tilde{\nu}$ state. In particular, the $\tilde{\nu}=1$ case produces the filled Landau level wavefunctions hierarchically, thus establishing the hierarchical nature of the integer quantum Hall states. It follows that the composite fermion description of fractional quantum Hall states fits within the hierarchy theory of the fractional quantum Hall effect. I also demonstrate this directly by generating the composite fermion ground-state wavefunctions via application of the hierarchy construction to fractional quantum Hall states, starting from the $\nu=1/m$ Laughlin states.
\end{abstract}

\maketitle











The discoveries of the integer quantum Hall (IQH)~\cite{von_Klitzing80} and fractional quantum Hall (FQH) effects~\cite{Tsui82}, where the Hall conductivity of a 2D electron gas subjected to low temperatures and large magnetic fields takes values $\sigma_{xy} = \nu e^2/h$ for integer and rational $\nu$, respectively, have led to an ongoing effort to understand their nature. The IQH effect can be understood in terms of independent electrons filling Landau levels, with the extremely precise quantization of the Hall conductance explained by gauge invariance and the existence of a mobility gap~\cite{Laughlin81} or, more rigorously, by a topological invariant called the Chern number~\cite{Thouless82,Bellissard94}.
Laughlin provided the phenomenological explanation for FQH states with filling $\nu=1/m$ (with $m$ odd) in terms of quantum fluids, for which he provided highly accurate trial wavefunctions~\cite{Laughlin83}. Laughlin's states possessed exotic new features, such as topological order~\cite{Wen90b} and quasiparticle excitations with fractional charge~\cite{Laughlin83} and braiding statistics~\cite{Arovas84}. Soon after Laughlin's proposal, Haldane and Halperin proposed the hierarchy theory~\cite{Haldane83,Halperin83,Halperin84} of similar states built on Laughlin states to explain the FQH effect at other odd-denominator filling fractions which were being discovered. Later, the composite fermion (CF) theory was proposed by Jain as an alternative description of these FQH states~\cite{Jain89}.

The CF description has been particularly successful because it provides a superficially simple physical picture of the FQH effect, a straightforward phenomenology that matches well with experiments, and trial wavefunctions which extensive numerical studies have found to be highly accurate and which permit certain computations for relatively large system sizes ($N \sim 100$); see, e.g.~\cite{Jain-book}, for a review. CF theory also provides a conceptual context for the $\nu=1/2$ Fermi-liquid-like state~\cite{Halperin93}. In contrast, the same range of phenomenology has not been as easy to elicit from the Haldane-Halperin (HH) hierarchy theory. Numerical studies of hierarchy generated wavefunctions are far less extensive, but have found them to also be highly accurate~\cite{Greiter94}. It has been argued that the HH hierarchy and CF states at the same $\nu$ are simply descriptions of the same universality class, since they possess identical topological quantum numbers, such as their quasiparticles' charge and braiding statistics~\cite{Read90}. In contrast, it has also been argued that the CF theory, which views the FQH states as IQH states of composite fermions, is fundamentally different from the hierarchy theory and that there is no hierarchical relation between different filling fractions in the CF theory, since there is none between different integer filled states (see, e.g. Ch.~12.1 of~\cite{Jain-book}). In Refs.~\cite{Hansson09a,Hansson09b,Suorsa11}, it was shown, working on the plane, that applying the hierarchical construction for specifically designed quasiparticle trial wavefunctions exactly produces CF wavefunctions. These quasiparticle wavefunctions, which are generated via conformal field theory operators, are equivalent to CF generated quasiparticle wavefunctions, up to ambiguities and freedoms associated with defining wavefunctions of both approaches on the plane~\cite{Hansson07a,Hansson07b}.

In this letter, I consider hierarchical constructions for quantum Hall systems on the (2D) sphere~\cite{Haldane83}, where the relevant wavefunctions are more precisely defined than on the plane. I apply the hierarchy theory to the $\nu = n$ IQH states~\footnote{In this letter, I assume that the electrons' spins are fully polarized and can be ignored. However, it is straightforward to generalize the analysis to the case with spin.} to produce $\nu = n+ \tilde{\nu}$ states, where $0 < \tilde{\nu} \leq 1$ is the filling of a quantum Hall state into which the quasielectrons are projected. The resulting $\nu = n+ \tilde{\nu}$ wavefunctions are exactly those of states with Landau levels $0, 1, \ldots, n-1$ filled and the filling $\tilde{\nu}$ quantum Hall state in the $n$th Landau level. This is used to explicitly demonstrate the hierarchical nature of the IQH effect and of the CF theory of the FQH effect. In the supplementary material (following the letter), I provide details to repeat this analysis for planar systems and address associated issues.

In the hierarchy theory~\cite{Haldane83,Halperin83,Halperin84}, one starts from a wavefunction $\Psi\left( {\bf R}_{\mu} ; {\bf r}_i \right)$ for a state of $N$ electrons with coordinates ${\bf r}_1 ,\ldots , {\bf r}_N$ that has $N_{\text{qp}}$ quasiparticles at positions ${\bf R}_{1},\ldots,{\bf R}_{N_{\text{qp}}}$. Then the quasiparticles are projected onto a quantum Hall type state with (pseudo-)wavefunction $\Phi\left({\bf R}_{\mu}\right)$ (by taking the inner product with respect to the quasiparticles' coordinates), giving the wavefunction for a new electronic state
\begin{equation}
\label{eq:integral}
\Psi^{\prime} \left( {\bf r}_i \right) = \int \prod_{\mu=1}^{N_{\text{qp}}} d^{2}{\bf R}_{\mu} \,\, \overline{\Phi \left( {\bf R}_{\mu} \right) } \, \, \Psi\left( {\bf R}_{\mu} ; {\bf r}_i \right) .
\end{equation}
The wavefunction $\Phi$ must be suitably chosen so that the integrand is single-valued and hence well-defined. This hierarchy construction can be associated with the physical picture where an additional quantum Hall fluid is introduced and coupled to the original one by bose condensing composites of quasiparticles (which, in general, are anyonic) from the two Hall fluids that form electrically neutral bosons. Such a topological bose condensation produces a new topological phase from the previous two~\cite{Bais09}. Requiring that a neutral boson composite of quasiparticles can be formed provides precisely the same restriction on the allowed Hall fluid that may be added and hence its associated wavefunctions $\Phi$, as does the single-valued integrand requirement.

The electron positions on the sphere are expressed using spinor coordinates ${\bf r}_{i} = \left( u_i , v_i \right)$, which are related to spherical coordinates $\left( \theta_i , \varphi_i \right)$ by
\begin{equation}
u_i = \cos\left(\theta_i / 2 \right) e^{i \varphi_i /2} , \qquad v_i = \sin \left( \theta_i / 2 \right) e^{- i \varphi_i /2}
.
\end{equation}
In spherical quantum Hall systems, a radial magnetic field ${\bf B} = \frac{\hbar N_{\phi} }{2 e R^2} \hat{\bf r}$ is produced by a magnetic monopole at the center of the sphere, corresponding to $N_{\phi}$ flux quanta passing through the surface (in units of $\phi_0 = h/e$). The number of flux is related to the filling fraction $\nu$ by $N_{\phi} = \nu^{-1} N - S$, where the shift $S$ is a topological quantum number (for clean systems) characteristic of how the electrons couple to spatial curvature in different quantum Hall states~\cite{Wen92c}. There are $N_j = N_{\phi}  + 2j+ 1$ orbitals in the $j$th LL given by the monopole harmonics~\cite{Wu76}, which form a complete, orthogonal basis. In spinor coordinates, these (single particle) orthonormal basis states are
\begin{equation}
\psi^{(j)}_{s,m} \left( {\bf r } \right) = \sqrt{\frac{\left(2s +1\right) \left(2s - j\right)! } { 4 \pi \left(s + m \right)! \left(s - m \right)! j!} } \,\, \mathcal{L}^{j} \left\{ u^{s + m} v^{s - m} \right\}
\end{equation}
where $j=0,1,\ldots$ specifies the LL, $s = \frac{1}{2} N_{\phi} + j$ corresponds to the eigenvalue $s \left(s + 1 \right)$ of total angular momentum ${\bf L}^2$ for orbitals in the $j$th LL, $m = -s, -s +1 , \ldots, s$ is the eigenvalue of $L^{z}$ (specifying the orbital within the $j$th LL), and $\mathcal{L} = \bar{v} \frac{\partial}{\partial u} - \bar{u}  \frac{\partial}{\partial v}$ is the LL raising operator.

Thus, a system filling the $n$ lowest Landau levels ($j=0,1,\ldots,n-1$) has $N_{\phi} = \frac{1}{n} N - n$ and is described by the (unnormalized) wavefunction
\begin{equation}
\chi_{n} \left( {\bf r}_i \right) = \sum_{\sigma \in S_N} \frac{\left( -1 \right)^{\sigma}}{N !} \,\, \prod_{j=0}^{n-1} \chi_{1}^{(j)} \left( {\bf r}_{\sigma \left( \mathcal{N}_j + a_j \right)} \right)
\end{equation}
where $ \mathcal{N}_j =  \sum_{r=0}^{j-1} N_r$, and $\chi_{1}^{(j)}$ is the wavefunction for a filled $j$th LL, given by
\begin{equation}
\chi_{1}^{(j)} \left( {\bf r}_{a} \right) = \sum_{\tau \in S_{N_j}} \frac{\left( -1 \right)^{\tau}}{N_j !} \, \prod_{a=1}^{N_j } \psi^{(j)}_{s_j, s_j - a + 1} \left( {\bf r }_{\tau \left( a \right)} \right)
,
\end{equation}
where $s_j = \frac{1}{2} N_{\phi} + j$.

In order to introduce $N_{\text{qe}}$ quasielectrons in the $\nu=n$ state (for which such localized point-like excitations have charge $-e$ and fermionic statistics), one decreases the flux by $1/n$ for each quasielectron to $N_{\phi} = \frac{1}{n} N - n - \frac{1}{n} N_{\text{qe}}$. This decreases the number of orbitals in each LL and forces electrons to occupy LLs above the $(n-1)$th. To form minimal uncertainty excitations that are maximally localized at specific points, one uses coherent states. Thus, the minimal energy quasielectrons are coherent states in the $n$th LL. The coherent state $\phi_{ s_n , {\bf R} }^{\left(n\right)} \left( {\bf r } \right)$ in the $n$th LL on the sphere, localized at the point ${\bf R} = \left( \alpha, \beta \right)$ (in spinor coordinates) is obtained by applying $\mathcal{L}^{n}$ to the $0$th LL coherent state~\cite{Haldane83,Greiter11}, which (with a preferred normalization) yields
\begin{eqnarray}
\label{eq:coherent}
\phi_{ s_n , {\bf R} }^{\left(n\right)} \left( {\bf r } \right) &=& \frac{2s_n +1}{4 \pi} \sqrt{\frac{\left( 2s_n - n \right)!}{\left(2s_n\right)! n!} } \, \mathcal{L}^{n} \left\{ \left( \bar{\alpha} u + \bar{\beta} v \right)^{2 s_n} \right\} \notag \\
&=& \sum_{m = -s_n}^{s_n} \overline{ \psi^{(0)}_{s_n,m} \left( {\bf R } \right)} \,\, \psi^{(n)}_{s_n,m} \left( {\bf r } \right)
.
\end{eqnarray}
Thus, the $N$ electron wavefunction for this state with $N_{\text{qe}}$ quasielectrons localized at positions ${\bf R} _{\mu} = \left( \alpha_\mu, \beta_\mu \right)$, where $\mu = 1,\ldots,N_{\text{qe}}$, is given (for $N_{\text{qe}} \leq N_n$) by
\begin{eqnarray}
\Psi_{n} \left( {\bf R}_\mu ; {\bf r}_i \right) &=&
\sum_{\sigma \in S_N} \frac{\left( -1 \right)^{\sigma}}{N !} \prod_{j=0}^{n-1} \chi_{1}^{(j)} \left( {\bf r}_{\sigma \left( \mathcal{N}_j + a_j \right)} \right) \notag \\
&& \qquad \quad \times \prod_{\mu = 1}^{N_{\text{qe}}} \phi_{ s_n , {\bf R}_{\mu} }^{\left(n\right)} \left( {\bf r }_{\sigma \left( \mathcal{N}_n + \mu \right) } \right)
\label{eq:Psi_qe}
\end{eqnarray}
where this uses new values of $N_j$ and $s_j$ corresponding to $N_{\phi} = \frac{1}{n} N - n - \frac{1}{n} N_{\text{qe}}$.

The next step in the hierarchy construction is projecting the quasielectrons of this wavefunction into another quantum Hall type wavefunction $\Phi\left( {\bf R}_{\mu} \right)$ for the quasielectrons, as in Eq.~(\ref{eq:integral}). Using $\overline {\Phi\left( {\bf R}_{\mu} \right)} = \tilde{\Psi}_{\tilde{\nu}} \left( {\bf R}_{\mu} \right)$, a $0$th LL state with filling $0 < \tilde{\nu} \leq 1$ and shift $\tilde{S}$, the integral can only be non-zero if $s_{n} = \tilde{s}_{0}$, i.e. $N_\phi +2n = \tilde{\nu}^{-1} N_{\text{qe}} - \tilde{S}$. This requires $N_{\text{qe}} = \frac{\tilde{\nu}}{n+\tilde{\nu}}N + \frac{n \tilde{\nu} \left( n + \tilde{S} \right)}{n+ \tilde{\nu}}$ and results in $N_{\phi} = \frac{1}{n+\tilde{\nu}} N - \frac{ n^2 + 2 n \tilde{\nu} + \tilde{\nu} \tilde{S}}{ n +\tilde{\nu}}$ for the new state.

By expanding $\tilde{\Psi}_{\tilde{\nu}}$ in terms of orbital occupation
\begin{equation}
\label{eq:Psi_tnu}
\tilde{\Psi}_{\tilde{\nu}} \left( {\bf R}_{\mu} \right) = \sum_{m_1 , \ldots , m_{N_{\text{qe}}} =-\tilde{s}_0}^{ \tilde{s}_0} C_{\left[  m_1 , \ldots , m_{N_{\text{qe}}} \right]} \prod_{\mu =1 }^{N_{\text{qe}}}  \psi^{(0)}_{\tilde{s}_0,m_{\mu}} \left( {\bf R }_\mu \right)
\end{equation}
where $C_{\left[  m_1 , \ldots , m_{N_{\text{qe}}} \right]}$ are the expansion coefficients, the integrals can be evaluated using the property
\begin{equation}
\int d^2 {\bf R } \,\, \psi^{(0)}_{s_n,m} \left( {\bf R } \right) \, \phi_{ s_n , {\bf R} }^{\left(n\right)} \left( {\bf r } \right)
= \psi^{(n)}_{s_n,m} \left( {\bf r } \right)
.
\end{equation}

The wavefunction resulting from using Eqs.~(\ref{eq:Psi_qe}) and (\ref{eq:Psi_tnu}) in the hierarchical construction for the new state is
\begin{eqnarray}
&& \!\!\!\!\!\!\!\!\!\!\!\! \Psi^{\prime}_{n+ \tilde{\nu} } \left( {\bf r}_i \right) =
\int \prod_{\mu=1}^{N_{\text{qe}}} d^2 {\bf R }_\mu \,\, \tilde{\Psi}_{\tilde{\nu}} \left( {\bf R}_\mu \right) \Psi_{n} \left( {\bf R}_\mu ; {\bf r}_i \right) \notag \\
&& \!\!\!\!\!\!\!\!\!\!\!\! = \sum_{\sigma \in S_N} \frac{\left( -1 \right)^{\sigma}}{N !} \prod_{j=0}^{n-1} \chi_{1}^{(j)} \left( {\bf r}_{\sigma \left( \mathcal{N}_j + a_j \right)} \right) \, \tilde{\Psi}_{\tilde{\nu}}^{(n)} \left( {\bf r }_{\sigma \left( \mathcal{N}_n + \mu \right)} \right)
\end{eqnarray}
where
\begin{equation}
\tilde{\Psi}_{\tilde{\nu}}^{(n)} \left( {\bf r }_{\mu} \right) = \sum_{m_1 , \ldots , m_{N_{\text{qe}}} =-s_n}^{ s_n} C_{\left[  m_1 , \ldots , m_{N_{\text{qe}}} \right]} \prod_{\mu =1 }^{N_{\text{qe}}}  \psi^{(n)}_{s_n,m_{\mu}} \left( {\bf r }_\mu \right)
\end{equation}
is the wavefunction obtained by raising $\tilde{\Psi}_{\tilde{\nu}}$ from the $0$th LL to the $n$th LL (and replacing $s_0$ with $s_n$). The wavefunction $\Psi^{\prime}_{n+ \tilde{\nu} } \left( {\bf r}_i \right)$ is exactly that of a $\nu=n+\tilde{\nu}$ state of $N$ electrons with Landau levels $0, 1, \ldots, n-1$ filled and $\tilde{\Psi}_{\tilde{\nu}}$ partially filling the $n$th Landau level.
It should also be clear that this mapping is invertible, i.e. the wavefunction $\tilde{\Psi}_{\tilde{\nu}}$ can be uniquely obtained from $\Psi^{\prime}_{n+ \tilde{\nu} }$.

Projecting the quasielectrons onto the (conjugate of) a filled LL, i.e. using $\Phi\left( {\bf R}_{\mu} \right) = \overline{ \chi_1 \left( {\bf R}_{\mu} \right) }$ in Eq.~(\ref{eq:integral}) of the hierarchy construction, produces $\Psi^{\prime}_{n+ 1 } \left( {\bf r}_i \right) = \chi_{n+ 1} \left( {\bf r}_i \right)$, the wavefunction for the state with the $n+1$ lowest LLs filled. Thus, I have explicitly demonstrated that the IQH states possess a hierarchical structure.

The CF theory generates $\nu = \frac{\nu^{\ast} }{ 2p\nu^{\ast} \pm 1}$ FQH states by viewing them as filling $\nu^{\ast}$ quantum Hall states of CFs~\cite{Jain89}, which are bound states of electrons with $\pm 2p$ quantized vortices. The associated ground-state trial wavefunction generated from that of a $\nu^{\ast}$ state with wavefunction $\Psi_{\nu^{\ast}}$ is~\footnote{A lowest LL projection may be applied to produce wavefunctions that are strictly in the $0$th LL.}
\begin{equation}
\label{eq:CF}
\Psi_{\nu} \left( {\bf r}_i \right) = \left[ \chi_{1} \left( {\bf r}_i \right) \right]^{2p} \Psi_{\pm \nu^{\ast}} \left( {\bf r}_i \right)
\end{equation}
where $\Psi_{-\nu^{\ast}}\left( {\bf r}_i \right) = \overline{\Psi_{\nu^{\ast}} \left( {\bf r}_i \right)}$ corresponds to using negative vortices. In particular, the experimentally prominent series of FQH states $\nu = \frac{n}{ 2pn \pm 1}$ corresponding to $\nu^{\ast} = n$ are thought of as IQH states of CFs. The other odd-denominator filling fractions are generated in the CF picture using $\nu^{\ast} = n + \tilde{\nu}$, where the filling fraction $0<\tilde{\nu}<1$ is that of a previously obtained state (or its particle-hole conjugate), e.g. $\nu^{\ast} = 4/3$ gives $\nu = \frac{ 4 }{ 8p \pm 3}$, which is relevant for the observed $\nu=4/11$ state~\cite{Pan03} (see Ch.~7.4 of~\cite{Jain-book} for more details). From this perspective, it is now obvious that the CF description of FQH states is hierarchical in nature, just as the IQH and $\nu = n + \tilde{\nu}$ FQH states are hierarchical in nature.

It is also worth examining the direct application of the hierarchy construction to the trial wavefunctions generated from the CF picture (rather than to the IQH states) in order to clarify the details of how the construction transfers over. Here, I will focus on the $\nu = \frac{n}{2pn + 1}$ states. I stress that the hierarchy picture applies to universality classes of quantum Hall states and is not immutably restricted to specific choices of their representative trial wavefunctions. One should, however, try to employ the most accurate trial wavefunctions $\Phi \left( {\bf R}_{\mu} \right) $ and $\Psi\left( {\bf R}_{\mu} ; {\bf r}_i \right)$ that one can produce and utilize for Eq.~(\ref{eq:integral}). The quasielectron trial wavefunctions generated by the CF picture are very accurate, so it is natural to use them in the hierarchical construction. The CF ansatz for the $\nu = \frac{n}{2pn + 1}$ states [starting from the $\nu=1/m = 1/(2p+1)$ Laughlin states at $n=1$, which have trial wavefunctions $\chi_1^m$] with quasielectrons is generated similar to Eq.~(\ref{eq:CF}), but using the wavefunction $\Psi_{n} \left( {\bf R}_\mu ; {\bf r}_i \right)$ of Eq.~(\ref{eq:Psi_qe}) for IQH states containing quasielectrons. This gives
\begin{equation}
\Psi_{\nu} \left( {\bf R}_\mu; {\bf r}_i \right) = \left[ \chi_{1} \left( {\bf r}_i \right) \right]^{2p} \Psi_{n} \left( {\bf R}_\mu ; {\bf r}_i \right)
.
\end{equation}
Before inserting this wavefunction into the hierarchy machinery, it is important to notice that, as written, its norm has nontrivial ${\bf R}_\mu$ dependence, i.e. it is not just a constant (up to corrections that are exponentially suppressed with the separation of quasielectrons). Additionally, explicit transformation of this wavefunction upon exchanging quasielectrons results in a factor of $-1$ corresponding to the fermionic statistics of quasiparticles in the $\Psi_{n}$ states, rather than the phase $e^{i \theta}$, where $\theta / \pi = 1 - \frac{2p}{2pn+1}$ is the braiding statistical angle of quasielectrons in the $\Psi_{\nu}$ state. (The correct braiding statistics is recovered once the Berry's phase contribution is included.) The properly normalized wavefunction that explicitly exhibits the braiding statistics is
\begin{equation}
\hat{\Psi}_{\nu} \left( {\bf R}_{\mu}; {\bf r}_i \right) = \left[ \frac{ \chi_{1} \left( {\bf R}_{\mu} \right)}{\left| \chi_{1} \left( {\bf R}_{\mu} \right) \right| } \right]^{\frac{ \theta }{\pi} - 1}
\frac{ \Psi_{\nu} \left( {\bf R}_\mu; {\bf r}_i \right) }{ \left\| \Psi_{\nu} \left( {\bf R}_\mu; {\bf r}_i \right) \right\|}
.
\end{equation}

If one uses $\Psi_{\nu} \left( {\bf R}_\mu; {\bf r}_i \right)$ in the hierarchical construction of Eq.~(\ref{eq:integral}) and projects the quasielectrons into a Laughlin type quantum Hall state, then the natural (though perhaps na\"ive) choice for the wavefunction into which the quasielectrons are projected would be $\Phi \left( {\bf R}_{\mu} \right) = \overline{ \left[ \chi_{1} \left( {\bf R}_{\mu} \right)\right]}^{2k - 1}$, for $k$ a positive integer. This exactly reproduces the CF generated FQH trial wavefunctions (with $k=1$ giving the $\nu = \frac{n}{2pn + 1}$ states) as it reduces to multiplying $\left[ \chi_{1} \left( {\bf r}_i \right) \right]^{2p}$ by the $\nu = n+ \tilde{\nu}$ wavefunction $\Psi^{\prime}_{n+ \tilde{\nu} } \left( {\bf r}_i \right)$ with $\tilde{\nu} = \frac{1}{2k-1}$. In this way, one can hierarchically generate all the CF ground-state trial wavefunctions, starting from the $\nu = 1/m$ Laughlin states' trial wavefunctions.

However, it may seem more appropriate to use $\hat{\Psi}_{\nu} \left( {\bf R}_{\mu}; {\bf r}_i \right)$ in the hierarchical construction, since it explicitly manifests the physical correlation and statistics properties of the quasielectrons. In this case, the natural choice for projecting into a Laughlin type state is $\hat{\Phi} \left( {\bf R}_{\mu} \right) = \overline{ \left[ \chi_{1} \left( {\bf R}_{\mu} \right)\right]}^{2k - \frac{ \theta }{\pi}}$. The detailed wavefunctions resulting from these two constructions are generally different. To produce exactly the same wavefunctions, one must either multiply $\hat{\Phi}$ or divide $\Phi$ by the factor $f \left( {\bf R}_{\mu} \right) = \left| \chi_{1} \left( {\bf R}_{\mu} \right) \right|^{\frac{\theta }{\pi} -1} \left\| \Psi_{\nu} \left( {\bf R}_\mu; {\bf r}_i \right) \right\|$.

For the sake of producing wavefunctions, one can introduce factors like $f \left( {\bf R}_{\mu} \right)$ into Eq.~(\ref{eq:integral}) at will, since it leaves the integral well-defined. In fact, one can view the inclusion of such factors as variational parameters of the trial wavefunctions~\cite{Girvin84b,Suorsa11}. However, if one demands strict adherence to the physical hierarchy picture with $\Psi$ and $\Phi$ accurately describing quantum Hall states, then one must be more careful and justify that such factors do not alter the universal properties of the wavefunctions. This is a very credible claim for the factors $f \left( {\bf R}_{\mu} \right)$ needed to exactly equate the detailed forms of the above wavefunctions. In particular, from the similarities with the analogous quasihole wavefunctions, one expects $\left\| \Psi_{\nu} \left( {\bf R}_\mu; {\bf r}_i \right) \right\| = \left| \chi_{1} \left( {\bf R}_{\mu} \right) \right|^{1 - \frac{\theta }{\pi}}$ up to corrections that are exponentially suppressed with the separation of quasielectrons. This property can be demonstrated for quasiholes in the Laughlin and the quasihole-based HH hierarchy states using the plasma analogy~\cite{Laughlin83,Bonderson11}. Assuming this expected property for quasielectrons, it follows that $f \left( {\bf R}_{\mu} \right) \approx 1$ except possibly when the quasielectrons are near each other, which is precisely where the relevant wavefunctions vanish and hence have negligible contribution to the integrals. Consequently, including the factor $f \left( {\bf R}_{\mu} \right)$ in the hierarchical construction should not significantly modify the resulting wavefunctions nor alter their universal properties. Refs.~\cite{Hansson09a,Hansson09b,Suorsa11} tacitly use such arguments in their hierarchical construction of the CF wavefunctions.

In summary, I have demonstrated that the IQH and $\nu = n + \tilde{\nu}$ FQH states are hierarchical in nature. Through its fundamental relation to these states, the CF description of FQH states shares the same hierarchical nature. I also demonstrated the hierarchical structure of the CF FQH states directly through application of the hierarchy construction to Laughlin and CF trial wavefunctions, using quasielectron wavefunctions generated from the CF picture. As such, one may view the CF descriptions of FQH states as particularly successful representatives of the HH hierarchy theory. One can now obtain the sought after phenomenology of the HH hierarchy theory for FQH states by simply adopting that of the CF theory.

It is also possible to construct hierarchies over non-Abelian quantum Hall states~\cite{Bonderson08}. When these are formed starting with Laughlin type quasiparticles (i.e. flux $\pm 1$ excitations that occur purely in the charge sector) in these states, they also admit a description by CF type wavefunctions~\cite{Bonderson08,Bonderson09a,Bonderson10b}. The results of this letter also apply to these cases.

\begin{acknowledgments}
I thank B.~A. Bernevig, T.~H. Hansson, J.~K. Jain, C.~V. Nayak, and especially J.~K. Slingerland for illuminating discussions and comments. I thank the Aspen Center for Physics for hospitality and support under the NSF Grant No. $1066293$.
\end{acknowledgments}


\begin{widetext}

\end{widetext}

\appendix

\section{Supplementary Material}

The analysis of this letter can also be carried out for the planar disk geometry. The orthonormal basis states in the plane are
\begin{equation}
\psi^{(j)}_{m-j} \left( {\bf r } \right) = \frac{1}{\sqrt {2 \pi 2^m m! j!} } \mathcal{L}^{j} \left\{ z^{m} e^{- \frac{1}{4} \left| z \right|^2 } \right\}
,
\end{equation}
where $j = 0,1, \ldots$ specifies the LL, $m = 0 , 1 , \ldots$ specifies the orbital with $L^{z}$ eigenvalue $m-j$, and planar coordinates are expressed as complex variables $z = x+iy$. The LL raising operator is $\mathcal{L} = \left( \frac{1}{2} \bar{z} - 2 \frac{\partial}{\partial z} \right)/\sqrt{2}$.

The (unnormalized) wavefunction for a system filling the $n$ lowest Landau levels ($j=0,1,\ldots,n-1$) for a disk-like region of the plane takes the form
\begin{equation}
\chi_{n} \left( {\bf r}_i \right) = \sum_{\sigma \in S_N} \frac{\left( -1 \right)^{\sigma}}{N !} \,\, \prod_{j=0}^{n-1} \chi_{1}^{(j)} \left( {\bf r}_{\sigma \left( \mathcal{N}_j + a_j \right)} \right)
\end{equation}
where $ \mathcal{N}_j =  \sum_{r=0}^{j-1} N_r$, and $\chi_{1}^{(j)}$ is the wavefunction for a filled $j$th LL, given by
\begin{equation}
\chi_{1}^{(j)} \left( {\bf r}_{a} \right) = \sum_{\tau \in S_{N_j}} \frac{\left( -1 \right)^{\tau}}{N_j !} \, \prod_{a=1}^{N_j } \psi^{(j)}_{a -j -1} \left( {\bf r }_{\tau \left( a \right)} \right)
.
\end{equation}
In this case, the numbers $N_j$ of occupied orbitals in the $j$th LL are not uniquely determined, as they were for spherical systems, so there is some freedom in distributing the electrons among the different LLs. (The total number of orbitals in each LL is, of course, infinite on the plane.) The $N_j$ should be chosen to give roughly the same radius of occupied states in each occupied LL, defining a disk-like region of quantum Hall fluid. For $n$ ``filled'' LLs, this requires the occupation to be roughly the same in each occupied LL and a standard choice is the equal distribution $N_j = N/n$ (for $j=0,1,\ldots,n-1$). The effect of different distributions $N_j$ is evident at the edge of the disk.

Coherent states (with a preferred normalization) in the $n$th LL on the plane are given by
\begin{eqnarray}
\phi_{ {\bf R} }^{\left(n\right)} \left( {\bf r } \right) &=& \frac{1}{2 \pi \sqrt{n!}} \mathcal{L}^{n} \left\{ e^{-\frac{1}{4} \left( \left| \eta \right|^2 + \left| z \right|^2 - 2 \bar{\eta} z \right)} \right\} \notag \\
&=& \frac{1}{2 \pi \sqrt{2^n n!}} \left( \bar{z} - \bar{\eta} \right)^n e^{-\frac{1}{4} \left( \left| \eta \right|^2 + \left| z \right|^2 - 2 \bar{\eta} z \right)} \notag \\
&=& \sum_{m = 0}^{\infty} \overline{ \psi^{(0)}_{m} \left( {\bf R } \right)} \,\, \psi^{(n)}_{m-n} \left( {\bf r } \right)
,
\end{eqnarray}
where $\eta = X + i Y$ expresses the coordinate ${\bf R}$ as a complex variable.

In order to introduce quasielectrons in a $\nu=n$ state, one uses $n$th LL coherent states. The $N$ electron wavefunction for such a state with $N_{\text{qe}}$ quasielectrons localized at positions ${\bf R} _{\mu}$, where $\mu = 1,\ldots,N_{\text{qe}}$, is given (for $N_{\text{qe}} \lesssim N_j$) by
\begin{eqnarray}
\Psi_{n} \left( {\bf R}_\mu ; {\bf r}_i \right) &=&
\sum_{\sigma \in S_N} \frac{\left( -1 \right)^{\sigma}}{N !} \prod_{j=0}^{n-1} \chi_{1}^{(j)} \left( {\bf r}_{\sigma \left( \mathcal{N}_j + a_j \right)} \right) \notag \\
&& \qquad \quad \times \prod_{\mu = 1}^{N_{\text{qe}}} \phi_{ {\bf R}_{\mu} }^{\left(n\right)} \left( {\bf r }_{\sigma \left( \mathcal{N}_n + \mu \right) } \right)
\end{eqnarray}
where $N = \sum\limits_{j=0}^{n} N_j$ and $N_n = N_{\text{qe}}$. This can be done by adding $N_{\text{qe}}$ electrons to a $\nu=n$ state (leaving the $N_j$ fixed and changing $N$), by moving $N_{\text{qe}}$ electrons from the various $j<n$ LLs into the $n$th LL (leaving $N$ fixed and changing the $N_j$), or by changing both $N$ and the $N_j$.

With these definitions and the expansion of an arbitrary filling $0<\tilde{\nu} \leq 1$ state $\tilde{\Psi}_{\tilde{\nu}}$ in terms of orbital occupation
\begin{equation}
\tilde{\Psi}_{\tilde{\nu}} \left( {\bf R}_{\mu} \right) = \sum_{m_1 , \ldots , m_{N_{\text{qe}}} =0 }^{ \infty } C_{\left[  m_1 , \ldots , m_{N_{\text{qe}}} \right]} \prod_{\mu =1 }^{N_{\text{qe}}}  \psi^{(0)}_{m_{\mu}} \left( {\bf R }_\mu \right)
,
\label{eq:plane_Psi_tnu}
\end{equation}
where $C_{\left[  m_1 , \ldots , m_{N_{\text{qe}}} \right]}$ are the expansion coefficients, the same evaluation of integrals can be carried out as before. The wavefunction resulting from the hierarchical construction for the new state (with $N$ electrons) is again
\begin{eqnarray}
&& \!\!\!\!\!\!\!\!\!\!\!\! \Psi^{\prime}_{n+ \tilde{\nu} } \left( {\bf r}_i \right) =
\int \prod_{\mu=1}^{N_{\text{qe}}} d^2 {\bf R }_\mu \,\, \tilde{\Psi}_{\tilde{\nu}} \left( {\bf R}_\mu \right) \Psi_{n} \left( {\bf R}_\mu ; {\bf r}_i \right) \notag \\
&& \!\!\!\!\!\!\!\!\!\!\!\! = \sum_{\sigma \in S_N} \frac{\left( -1 \right)^{\sigma}}{N !} \prod_{j=0}^{n-1} \chi_{1}^{(j)} \left( {\bf r}_{\sigma \left( \mathcal{N}_j + a_j \right)} \right) \, \tilde{\Psi}_{\tilde{\nu}}^{(n)} \left( {\bf r }_{\sigma \left( \mathcal{N}_n + \mu \right)} \right)
\label{eq:plane_Psi'}
\end{eqnarray}
where
\begin{equation}
\tilde{\Psi}_{\tilde{\nu}}^{(n)} \left( {\bf r}_{\mu} \right) = \sum_{m_1 , \ldots , m_{N_{\text{qe}}} =0 }^{ \infty } C_{\left[  m_1 , \ldots , m_{N_{\text{qe}}} \right]} \prod_{\mu =1 }^{N_{\text{qe}}}  \psi^{(n)}_{n-m_{\mu}} \left( {\bf r }_\mu \right)
\label{eq:plane_Psin_tnu}
\end{equation}
is the wavefunction obtained by raising $\tilde{\Psi}_{\tilde{\nu}}$ from the $0$th LL to the $n$th LL.

Working with a finite number of electrons $N$ in the infinite plane introduces another potential issue. Since orbitals $\psi^{(j)}_{m-j}$ are only filled for $m < N_j$ (for $j<n$), producing a finite disk-like region, it may seem appropriate to similarly restrict the coherent states to orbitals with $m<M$ for some $M \sim N_j$, so that the support of the quasielectrons is the same as that of the quantum Hall droplet. (On the other hand, one might argue that such a truncation is unnecessary as long as the quasiparticle coordinates ${\bf R }_\mu$ remain ``inside'' the area of the disk of quantum Hall fluid, since then the coherent states will have Gaussian suppressed support in the orbitals with $m \geq M$, which are considered to be ``outside'' the disk.)

A truncated $n$th LL coherent state written as
\begin{equation}
\phi_{ {\bf R} }^{\text{tr} \left(n\right)} \left( {\bf r } \right) = \sum_{m = 0}^{M-1} \overline{ \psi^{(0)}_{m} \left( {\bf R } \right)} \,\, \psi^{(n)}_{m-n} \left( {\bf r } \right)
\end{equation}
has an ${\bf R}$-dependent norm
\begin{equation}
\left\| \phi_{ {\bf R} }^{\text{tr} \left(n\right)} \left( {\bf r } \right) \right\|^2 = \frac{1}{2 \pi } \frac{ \Gamma\left( M , \frac{\left| {\bf R} \right|^2}{2} \right) }{\Gamma\left( M \right) }
,
\end{equation}
in contrast with $\left\| \phi_{ s_n , {\bf R} }^{\left(n\right)} \left( {\bf r } \right) \right\|^2 = \frac{2s_n + 1}{4\pi}$ for the sphere and $\left\| \phi_{ {\bf R} }^{\left(n\right)} \left( {\bf r } \right) \right\|^2 = \frac{1}{2\pi}$ for the plane. Consequently, if truncated coherent states are used to describe quasielectrons and the properly normalized wavefunctions are used in the hierarchy construction, there will be additional factors of $\left[\Gamma\left( M , \frac{\left| {\bf R} \right|^2}{2} \right)\right]^{-1/2}$ in the integrands.

The effect of these normalization factors is that the expansion coefficients of $\tilde{\Psi}_{\tilde{\nu}}$ are modified by the hierarchy step to become
\begin{equation}
C^{\text{tr}}_{\left[  m_1 , \ldots , m_{N_{\text{qe}}} \right]} = C_{\left[  m_1 , \ldots , m_{N_{\text{qe}}} \right]} \prod_{\mu =1 }^{N_{\text{qe}}} g_{m_{\mu}} (M)
,
\end{equation}
where
\begin{equation}
g_{m} (M) \equiv \int d^2 {\bf R } \,\, \left| \psi^{(0)}_{m} \left( {\bf R } \right) \right|^2 \left[\frac{\Gamma\left( M \right) }{ \Gamma\left( M , \frac{\left| {\bf R} \right|^2}{2} \right) }\right]^{\frac{1}{2}}
.
\end{equation}
Thus, the $n$th LL component of the resulting hierarchical wavefunction becomes
\begin{equation}
\tilde{\Psi}_{\tilde{\nu}}^{\text{tr}(n)} \left( {\bf r}_{\mu} \right) = \sum_{m_1 , \ldots , m_{N_{\text{qe}}} =0 }^{ \infty } C^{\text{tr}}_{\left[  m_1 , \ldots , m_{N_{\text{qe}}} \right]} \prod_{\mu =1 }^{N_{\text{qe}}} \psi^{(n)}_{n-m_{\mu}} \left( {\bf r }_\mu \right)
,
\end{equation}
which is used in Eq.~(\ref{eq:plane_Psi'}) instead of Eq.~(\ref{eq:plane_Psin_tnu}). This modification should not alter the universal properties of the state, at least for sufficiently large $M$. In fact, taking the thermodynamic limit (where $N$, $N_j$, and $M$ all go to infinity) gives $g_{m} (M) \rightarrow 1$ and recovers the wavefunction of Eq.~(\ref{eq:plane_Psin_tnu}).

\end{document}